\documentclass[12pt,a4paper,final]{iopart}
\usepackage{graphicx}

\usepackage{iopams}  
\usepackage[breaklinks=true,colorlinks=true,linkcolor=blue,urlcolor=blue,citecolor=blue]{hyperref}
\usepackage[utf8]{inputenc}
\usepackage[english]{babel}
\expandafter\let\csname equation*\endcsname\relax
\expandafter\let\csname endequation*\endcsname\relax
\usepackage{amsmath}

\begin{document}

\title[]{Quantitative evaluation of laser-induced fluorescence in magnetized plasma accounting for disalignment effect}
\author{Roman Bergert, Slobodan Mitic, Markus H. Thoma}
\address{Institute of Experimental Physics I, Justus-Liebig-University Giessen Heinrich-Buff-Ring 16,
D-35392 Giessen, Germany}
\ead{Roman.Bergert@physik.uni-giessen.de}

\begin{abstract} Quantitative evaluation of tunable diode laser induced fluorescence (TDLIF) measurements in magnetized plasma take into account Zeeman splitting of energetic levels and intra-multiplet mixing defining the density distribution (alignment) of excited $\rm 2p_8$ multiplet is discussed in this paper. TDLIF measurements were used to evaluate light-transport properties in a strongly magnetized optically thick argon plasma under different pressure conditions. Therefore, a coupled system of rate balance equations were constructed to describe laser pumping of individual magnetic sub-levels of $\rm 2p_8$ state through frequency separated sub-transitions originating from $\rm 1s_4$ magnetic sub-levels. 

The density distribution of $\rm 2p_8$ multiplet was described by balancing laser pumping with losses including
radiative decay, transfer of excitation between the neighboring multiplets driven by neutral collisions and quenching due to electron and neutral collisions.
Resulting $\rm 2p_8$ magnetic sub-level densities were then used to model polarization dependent fluorescence, consider self-absorption, which could be directly compared with measured polarization resolved TDLIF measurements. This enables to obtain unique solutions for the $\rm 1s_4$ and $\rm 1s_5$ magnetic sub-level densities which were in good agreement with the densities obtained by laser absorption measurements. It is shown that LIF measurements in magnetized plasma conditions have strong pressure dependence that should be corrected consider effective disalignment rate. The presented measurement method and model can help further understanding and improve description of optical emission of argon in magnetized conditions.
\end{abstract}


\maketitle

\section{Introduction}
Influence of an external magnetic field $B$ changes the plasma behavior by influencing movement of a charged particles by the Lorentz force and changing energetic structures of a plasma components by the Zeeman effect \cite{zeeman1897xxxii} which overall influence the plasma properties. The presence of an external magnetic field in plasma induces anisotropy in motion of a charged particles by reducing cross-field diffusion, while also inducing strong anisotropy in plasma optical emission spectra. Total emitted line intensity is therefore artificial construct of two polarized components orientated parallel ($\pi$-component) or perpendicular ($\sigma\pm$-component) to the magnetic field vector. Due to such complex effects of magnetic field on plasma components, standard (non-magnetized) description of plasma emission spectra and light transport properties are not suitable.
The complexity of optical properties mainly comes from the splitting of energy levels. The total angular momentum quantum number $J$ of an energetic level will result in $2J+1$ magnetic sub-levels , each sub- level is described by the magnetic quantum number $m$, in the presence of a magnetic field compared to the one single level without a magnetic field. This changes the emission into a system of sub-transitions symmetrically redistributed around the initial unshifted line center. As a result, each spectral transition will have to be rewritten as a sum of optically allowed transitions between corresponding upper and lower magnetic sub-levels \cite{taylor2017zeeman}. 
The degeneracy of the levels vanishes under higher external magnetic fields so that an individual description of transitions between magnetic sub-levels is necessary \cite{fujimoto2008plasma}. In optically thick plasma, density distribution of $\rm 1s$ sub-states should be taken into account for correct description of the self-absorption effect and thus the light transport \cite{Bergert_2019}.

Description of polarized plasma spectroscopy has been introduced by Fujimoto and colleagues \cite{fujimoto2008plasma,Kazantsev1987,fujimoto1997plasma,GOTO2003} who presented a population-alignment collisional-radiative model (PACRM) based on collisional and radiative interactions between the multiplets and intra-multiplet transitions. Such approach has been demonstrated on the experiments where polarization of plasma optical emission spectroscopy (OES) was induced by highly anisotropic electron energy distribution function (electron beam), mostly in helium or hydrogen, while only limited number of reports considered argon. Also such description is highly non-trivial accounting for less available collisional cross-sections between different multiplets (ground, 1s and 2p states). The possibilities of PACRM for helium has been demonstrated in articles deducing highly interesting plasma properties, such as an anisotropy in the electron energy distribution function based on the modelling of the polarities of a plasma emission lines \cite{iwamae2005anisotropic,teramoto2018}.

The main goal of this work is to evaluate optical emission properties in magnetized argon plasma accounting for Zeeman splitting of radiating ($\rm 2p_8$) and $\rm 1s$ stats, density distribution between magnetic sub-levels (alignment) and polarization of plasma emission. Therefore, laser absorption spectroscopy (LAS) and laser induced fluorescence (LIF) were used as diagnostic tools. Description of LAS and evaluation of argon $\rm 1s_4$ and $\rm 1s_5$ sub-level state densities has been introduced in a recent article \cite{Bergert_2019}. Based on the sub-level densities the self-absorption (SA) for different sub-transitions was discussed. Self-absorption was included in the interpretation of linear polarized and circular polarized TDLIF at 842 nm and 801 nm. Optical sub-transitions between $\rm 1s_4$ and $\rm 2p_8$ sub-states were used to pump different magnetic sub-level of $\rm 2p_8$. 

A model describing the polarization dependent fluorescence was developed including self-absorption and sub-level mixing (disalignment) due to neutral collisions which strongly influence the ratio between the two polarities of the fluorescence spectral lines. Such effect changes the description of 2p state densities by introducing disalignment as additional production and losses mechanism competing with radiative decay. The correct analytical description will result in quite strong coupling between involved states densities through disalignment and self-absortion so that LIF measurements can be uniquely quantitatively solved resulting in estimations of the 1s magnetic sub-levels densities and disalignment constant. The disalignment constant has been estimated in \cite{Matsukuma2012} where in the presence of weak magnetic field, time dependent polarized LIF measurements has been modeled. However, in most of such experiments, broad and powerful lasers were used for inducing excitation followed by modeling of polarized emission. Another interesting aspect of disalignment constant is the temperature dependence which can cause quite opposite effect depending on the quenching atoms \cite{Matsukuma2012,Grandin1981}. In the reported case, krypton atoms reduced disaligment with temperature increase while for helium, neon and argon disalignment constant was increased. Such findings are illustrating the complexity of the effect and quite interesting non-linear response of different argon 2p levels. Although such unique optical properties for magnetized plasma conditions has been investigated since 1960's using highly complicated PACRM formalism, strong and clearly visible effect of neutral gas density on plasma spectra and its practical applications has been poorly conveyed to the plasma spectroscopy audience.
Therefore, we present a model for the quantitative evaluation of LIF measurements using tunable diode laser,  carried out on an compact dielectric barrier discharge low pressure plasma in the presence of an external magnetic field of 0.3 T.

\section{Methods}

Correct interpretation of optical transitions (absorption and induced fluorescence) in magnetized plasma conditions is based on line splitting formalism due to Zeeman effect (resulting in polarization) and its further implications in light transport properties under optically thick conditions. Direct detection of individual induced fluorescence profiles were done with a spectrometer, where a slow laser scan over a broad frequency range was used for pumping frequency shifted sub-transitions. Additionally, laser absorption spectroscopy of the $\rm 1s_4$ state was used as a reference for the sub-level densities evaluated from TDLIF model.

\begin{figure}[h]
	\centering
	\includegraphics[width=0.4\linewidth]{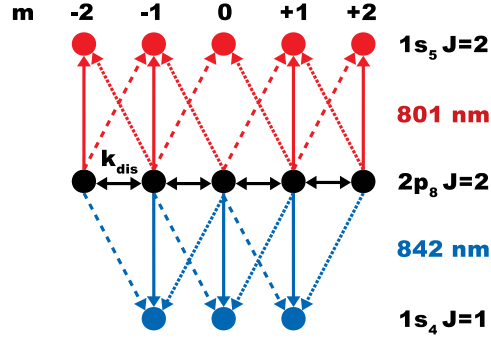}
	\caption{Kastler diagram indicating possible optical transitions under the influence of an external magnetic field. ($\pi$ solid lines, $\sigma -$ dashed lines and $\sigma +$ dotted lines) Additionally, the disalignment effect with the disalignment constant $k_{\rm dis}$ between the adjacent sub-levels of $\rm 2p_8$ is indicated with solid black arrows. }
	\label{fig:kastler}
\end{figure}

Under the influence of the magnetic field, each energetic level will split into the 2J+1 multiplet producing system of transitions described by the transition rules. In a strong magnetic field, sub-transitions will be shifted in frequency producing absorption structure symmetric around unshifted line center. The observed energetic structure and system of transitions of 801 and 842 nm in argon are represented by a Kastler diagram as shown in \ref{fig:kastler}. The allowed electric dipole transitions originating from $\rm 2p_8$ to $\rm 1s_4$ and $\rm 1s_5$ are shown by their polarization. The $\rm 2p_8$ and $\rm 1s_5$ state splits into 5 sub-levels. Hence, 12 different electric dipole transitions are formed (4 $\sigma+$, 4 $\pi$ and 4 $\sigma-$), which is indicated with red arrows between the sub-levels of $\rm 2p_8$ and $\rm 1s_5$. Nine sub-transitions can be found between $\rm 2p_8$ and $\rm 1s_4$ (3 $\sigma+$, 3 $\pi$ and 3 $\sigma-$), which is indicated with blue arrows in figure \ref{fig:kastler}. 

\subsection{Tunable diode laser absorption spectroscopy}
\label{subsec:absorption}

The description of LAS, polarization dependent sub-level transitions and resulting self-absorption has been described in our previous work \cite{Bergert_2019} and it will be used as a basis for LIF interpretation. Therefore, just a short overview will be given on LAS and description of magnetic sub-transitions rates since this has a direct impact on optical properties like self-absorption and light transport.
Based on the reconstructed absorption profile of individual sub-transition $\kappa(\nu)$ the magnetic sub-level density $n_m{_j}$ of targeted 1s states can be calculated according to:
\begin{equation}
    n_m{_j}=\frac{8\pi}{c^2}\frac{\nu_{m_i,m_j}^2}{A_{J_iJ_j,m_im_j}}\kappa(\nu)
\end{equation}
where $c$ describes the vacuum speed of light, $\nu_{m_i,m_j}$ is a central wavelength of the (shifted) transition dependent on the external magnetic field strength, and the Einstein coefficient of transition $A_{J_{i},J_{j},m_{i},m_{j}}$ between upper $\rm 2p_8$ $m_i$ and lower $\rm1s_4$ or $\rm 1s_5$ magnetic sub-level states $m_j$. The expression for the Einstein coefficients describing transitions between upper and lower magnetic sub-levels can be written as \cite{takacs1996polarization,jacobs1999angular}:
\begin{equation}%
\begin{split}
    A_{J_i,J_j,m_i,m_j}= \left(2J_i+1\right)A_{ij}\times W\\\
     W=\left|\left(\begin{matrix}
 J_j & q & J_i\\
 -m_j & \Delta m_{j,i} & m_i
 \end{matrix}\right)\right|^2
 \end{split}
\label{eq:Einstein} 
\end{equation}
where W represents the Wigner 3-j coefficient to the power of two with the multi-polarity of the transition $q$ equivalent to 1 for electrical dipole transitions, and difference in the magnetic quantum number of the transition $\Delta m_{j,i}= m_j-m_i$. $A_{ij}$ describes the Einstein coefficient (life time) for spontaneous decay of the observed emission in unmagnetized case.

\subsection{Tunable diode laser induced fluorescence in magnetized plasma}

The laser induced fluorescence measurements were based on the same laser excitation scheme consecutively pumping different $\rm 2p_8$ sub-levels through 842 nm sub-transitions separated by frequency. The efficiency of laser pumping is therefore proportional to the laser intensity, density of the targeted $\rm 1s_4$ $m_j$-multiplet and Einstein coefficient of sub-transition assuming a linear regime for laser pumping.
Each transition interacts only with corresponding polarity from the un-polarized pumping laser so that laser intensities for different polarization should scale as $I_\pi=2I_\sigma$. 

Pumping of an individual $\rm 2p_8$ sub-level would lead to fluorescence of the light emission with polarization determined by the ratio of Einstein coefficients of $\pi$ and $\sigma$ component originating from the pumped level and corrected for self-absorption. 

However under the observed conditions (magnetically separated transitions) it is evident that the polarization of the induced fluorescence is strongly effected by an additional process leading to appearance of an additional intensities, changing the expected polarization of the fluorescence, which in some cases could not be described by the transition rules. The observed deviations from the predicted polarization are described by inter-multiplet state density mixing driven by collision with neutrals. Such effect would try to equalize the density distribution within multiplets and destroy the alignment. The disalignment effect needs to be considered, along other mechanism, as an additional production and loss mechanism in the rate balance equation describing $\rm 2p_8$ sub-state densities.

 Measured intensities were corrected by subtracting plasma emission consider polarization dependency. Due to the optical selection rules there are no allowed electric dipole $\pi$ transitions for $m_i=2$ to $m_j=2$ of the $\rm 2p_8$ to $\rm 1s_4$ system and  $m_i=0$ to $m_j=0$ of the $\rm 2p_8$ to $\rm 1s_5$ system. Nevertheless, the signals were very strong when observing these systems especially when pumping $m_i=2$ and measuring $\pi$ fluorescence of 842 nm. These intensities raise with higher pressure indicating the role of level mixing by neutral atoms. Such effect would transfer population from laser pumped level ($m_j=2$) to the adjacent magnetic sub-levels resulting in polarization of light emission determined by final density distribution achieved among magnetic sub-levels. The resulting disalignment constants for different 2p states of argon are discussed in \cite{Grandin1981,grandin1978depolarisation,grandin1973sections}.

\section{Experimental Setup}

\begin{figure}[h]
	\centering
	\includegraphics[width=0.5\linewidth]{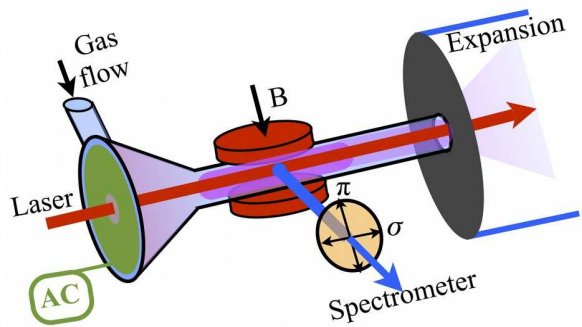}
	\caption{Schematic representation of the experimental setup.}
	\label{fig:set-up}
\end{figure}

The experimental setup and discharge configuration is presented in figure \ref{fig:set-up} with indicated laser path (red arrow) and line of sight for LIF detection (blue arrow). A cone shaped glass discharge chamber followed by a tube of 4 mm inner diameter were used to create a compact plasma jet. Open end of the tube was attached to 500 mm long expansion chamber with the vacuum system attached at the end. Driven electrode made of aluminum tape was attached from the outside of the cone base with an opening in the center to introduce a laser beam in axial direction.

A base pressure of 1.4$ \times 10^{-1}$ Pa could be reached. The minimum working pressure inside the tube was estimated to be around 46 Pa at 0.014 slpm gas flow and highest 431 Pa at 0.32 slpm. The working pressure inside the tube was estimated based on the measured Doppler shift in the gas velocity distribution function caused by the induced gas flow through the plasma jet similar as in \cite{kaupe2018phase,mitic2019comparative}. The particle velocity, extracted from Doppler shift measurements, can be correlated to the pressure inside the jet through the conservation of volume flow through the jet accounting for the cross-section of the tube and temperature of the gas. 
The Doppler shift was measured in fluorescence mode using axial laser excitation, along the gas flow, and radial fluorescence measurements at the point between the magnets. The fluorescence signal was detected by photo-diode and recorded together with Fabry-P\'{e}rot signal for scale and reference. The intensity of the shift was estimated by comparing with the unshifted position determined by radial laser absorption measurements.

Plasma was created by a bipolar 30 kHz sinusoidal high voltage ($V_{pp}\approx$ 4 kV) signal. The external magnetic field was introduced by two cylindrical permanent magnets of 6 mm in diameter inducing a magnetic field of 0.3 T. The field strength was confirmed by a gaussmeter.
The magnets were mounted perpendicular to the line of sight at the opposite sides at the discharge tube.

Laser absorption measurements were done in radial direction crossing the tube at position between the magnets, including common optical elements like argon reference cell and a Fabry-P\'{e}rot interferometer to monitor laser scanning range and quality. Absorption at 842.47 nm transition was scanned by \textit{TOPTICA DLC 100} laser. A system of collimators and multimode $200~\mathrm{\mu m}$ optical fiber was used to manipulate the laser beams, producing an unpolarized probing laser light. A band pass filters at 840 nm with a full width at half-maximum of 10 nm were used in order to suppress the rest of the plasma emission. The absorption measurements were done using optics crossing the tube in radial direction in orientation perpendicular to the magnetic field lines allowing to probe all polarization transitions.

For fluorescence measurements, laser light was introduced perpendicular to the external magnetic field in axial direction of the discharge tube along the gas flow. The 3 $\pi$ and 3 $\sigma+$ transitions of $1s_4$ at 842 nm were used to pump the $2p_8$ sub-levels. Laser scan was set at low scanning frequency allowing to measure simultaneously the induced fluorescence with a high sampling rate (10 ms integration time) using $200~\mathrm{\mu m}$ optical fiber and a collimator connected to Ocean Optics (USB2000+) spectrometer. The fluorescence was observed perpendicular to the illumination laser using identical optics from the LAS measurements.
A linear polarization filter from Thorlabs GmbH with a total transmission of 40 \% for unpolarized light was used between the plasma and collection optics to isolate desired polarization from the induced fluorescence. With linear polarization filter oriented parallel to the magnetic field lines only $\pi$ transitions could be transmitted while perpendicular orientation would transmit only $\sigma$. Polarization filter was not completely isolating desired polarity so that 3 \% transmissions of the opposite polarity was measured which was further taken into account for correct reconstruction of polarization dependent intensity. Relative sensitivity for linear and circular polarized light was evaluated to 1:0.7, which is usually dependent on an optics and detector system. Such difference in sensitivity of the two polarities was found to be between 0.581 to 0.67 in other studies \cite{Matsukuma2012,Seo_2003,Wakabayashi1998}.

\section{Results and Discussion}
\label{sec:resultsanddiscussion}

\begin{figure}[h]
	\centering
	\includegraphics[width=0.5\linewidth]{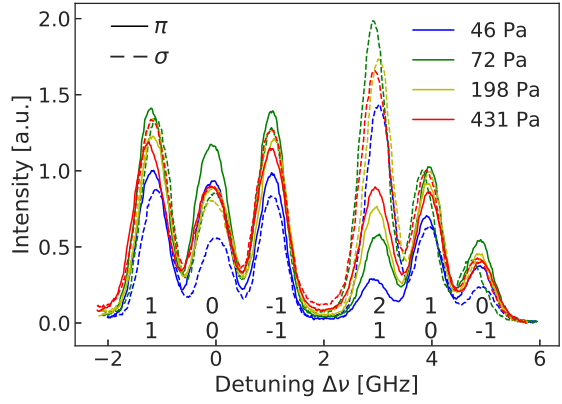}
	\caption{Pressure dependency of measured polarized fluorescence at 842 nm for four different pressures. The $A_{21}$ peak of the $\pi$ profile exhibits the crucial intensity increase due to the disalignment effect. Numbers in the lower part of the figure are indicating the upper $m_i$ and lower $m_j$ magnetic quantum number of each transition.}
	\label{fig:s4measurements}
\end{figure}

Fluorescence emission recorded at 842 nm branch for 4 different gas pressures are presented in figure \ref{fig:s4measurements} colour-coded with thick lines for $\pi$ and dashed for $\sigma$ component. The magnetic sub-level quantum number of the levels involved in the transitions are indicated at the bottom of the figure. The relative changes of the intensities with pressure are noticeable for each polarization component. The intensity of not allowed $\pi$ component originating at $A_{21}$ transition is clearly visible demonstrating the influence of the disalignment on the light transport properties in magnetized plasma. 

\begin{figure}[h]
	\centering
	\includegraphics[width=0.5\linewidth]{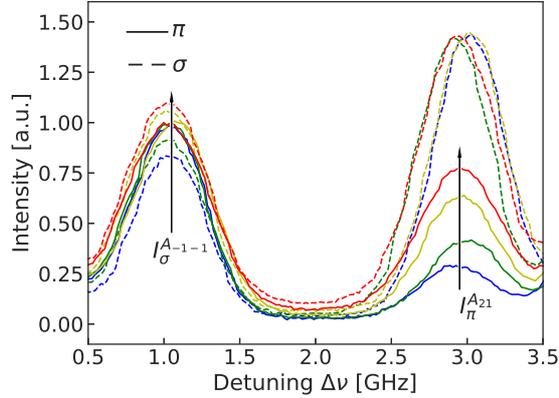}
	\caption{Systematic changes of measured 842 nm polarized fluorescence with pressure scaled to the $A_{-1-1}$ $\pi$ component. Strong increase of not optically allowed $A_{21}$ $\pi$ component is clearly increasing with pressure as a result of more efficient excitation transfer between the sub-states.}
	\label{fig:s4cut}
\end{figure}

Systematic changes in polarization of the fluorescence induced at $m_j=-1$ to $m_i=-1$ and $m_j=1$ to $m_i=2$ sub-transitions are better presented in figure \ref{fig:s4cut} with intensities scaled by the intensity of the respective $m_i=-1$ to $m_j=-1$ $\pi$ peak at each pressure.
From such representation it is evident that $\pi$ component only slightly changes the shape with pressure increase for all except $m_i=2$ to $m_j=1$  transition which should be pure sigma by the optical selection rules for electrical dipole transitions.

Considering the mentioned Kastler diagram (figure \ref{fig:kastler}) and possible transitions it is evident that excitation of different $\rm 2p_8$ multiplet ($m_i=-2,-1,...,2$) will produce unique ratio of $\pi$ to $\sigma$ fluorescence components for both 842 nm and 801 nm branches. Furthermore, optical pumping of $\rm 2p_8$ $m_i=\pm 2$ is only possible by $\sigma$ polarized component from $\rm 1s_4$ $m_j=\pm 1$ and according to the optical selection rules for electrical dipole transitions only $\sigma$ polarized light at 842 nm branch should be emitted. However, from the presented measurements the $\pi$ component detected by the excitation of $m_i=2$ sub-level state is clearly visible and rather strong with systematic increase with pressure. This is a solid evidence of the disalignment process or intra-multiplet density mixing driven by collisions with the neutral atoms. Such effect would transfer excitation from the laser pumped sub-level ($m_i=2$) to the neighboring magnetic sub-levels and vice-versa producing a light emission with polarization determined by the final density distribution achieved within the multiplet called alignment. Such effect is also responsible for detection of not optically allowed $\pi$ component on 801 nm branch when $\rm 2p_8$ $m_i=0$ state is optically pumped.

Measured fluorescence was first analyzed by identifying the positions and width of the transitions from measured fluorescence structure, quite similar to the LAS structure analyzed in our previous work \cite{Bergert_2019}. 
Laser induced fluorescence was modeled including collisional and radiative processes contribution to final state density distribution. 
In order to model measured fluorescence rate balance equation describing $\rm 2p_8$ sub-levels density induced by laser pumping was constructed including radiative decay, self-absorption, disalignment effect, stimulated emission and quenching effects. With slow laser scan, steady state solution for each of $\rm 2p_8$ multiplet can be described in following form: 

\begin{equation}
\begin{split}
  n_{{\rm 2p_8},m_i}&=\frac{I_{\pi/\sigma} B^{\rm 842}_{m_jm_i}n_{{\rm 1s_4},m_j}+N k_{\rm dis}\sum n_{{\rm 2p_8},{m_i\pm 1}}}{\sum\left(A^{\rm 842}_{{m_im_j}}\gamma_{m_i,m_j}(n_{{\rm 1s_4},m_j})+ A^{\rm 801}_{{m_im_j}}\gamma_{m_i,m_j}(n_{{\rm 1s_5},m_j})\right)+2N k_{\rm dis}+Q}\\ 
  Q&=I_{\pi/\sigma} B^{\rm 842}_{m_jm_i}\gamma_{m_i,m_j}(n_{{\rm 1s_4},m_j})+c_e+c_{q}
\end{split}
\label{eq:n2p}
\end{equation}

The $I_{\pi/\sigma}$ describes either the linear or circular polarized laser intensity, $N$ is the number of the neutral atoms, $B_{m_j,m_i}$ is the Einstein coefficient for absorption between magnetic sub-levels. The term under the sum in the denominator is describing the lifetime of the 2p sub-level including all the radiative transitions corrected by corresponding self-absorption factors $\gamma_{m_i,m_j}$. The self-absorption coefficient is dependent on the magnetic sub-level density $n_{1s,m_j}$ and the absorption coefficient. If a sub-state is not directly pumped than the first factor in numerator in equation \ref{eq:n2p} describing laser pumping should be neglected. The laser stimulated emission for the 842 nm branch is taken into account only for the pumped sub-states and is described by the term $I_{\pi/\sigma} B^{\rm 842}_{m_jm_i}\gamma_{m_i,m_j}(n_{{\rm 1s_4},m_j})$ in the denominator. Quenching of 2p states by collision with electrons $c_e$ \cite{PALOMARES2013156} and neutrals $c_q$ \cite{Chang1978,Nguyen1978} are also accounted as a loss mechanisms in the rate balance equation. Electron densities of few $10^{17}\, \rm m^{-3}$ were assumed to calculate the rate constant due to electron collision. The electron density is assumed to be similar to un-magnetized conditions from our previous work \cite{kaupe2018phase}.

Rate balance equation accounts the laser pumping as a source of population proportional to Einstein coefficient and targeted 1s sub-level density while additional production due to disalignment is included as transfer of excitation from neighboring sub-levels described with rate constant $k_{\rm dis}$ in $\rm m^3/s$. The disalignment effect is accounted as a loss mechanism, competing mostly with radiative decay, in depopulation of the described state. The term $2Nk_{\rm dis}$ in the denominator of the equation (\ref{eq:n2p}) losses the factor 2 in description of $m_i=\pm2$ states since transfer from only one neighboring 2p sub-level is possible.

The laser intensity in our experiments was measured to be 15 mW focused on the 2 mm diameter spot resulting in laser intensity of few hundreds $\rm W/m^2$ which according to \cite{Demtroder} is well above laser saturation intensity. With saturated laser intensity, the laser excitation is no longer dependent on the laser intensity. But secondary emission start to play a role and is taken into account.

The system of equations can be build for each probed transition providing laser-induced alignment of $\rm 2p_8$ state that can be further used to model the intensity of the fluorescence.
Solved sub-level densities were further used to describe emission rate of both polarities induced by laser pumping in a following form for 842 nm branch: 
\begin{equation}
\begin{split}
I^{\rm 842nm}_{\pi}&=n_{2p_8,0}A_{00}\gamma_{0,0}(n_{\rm 1s_4,0})+n_{\rm 2p_8,1}A_{11}\gamma_{1,1}(n_{\rm 1s_4,1})+n_{\rm 2p_8,-1}A_{-1-1}\gamma_{-1,-1}(n_{\rm 1s_4,-1})\\
\\
I^{\rm 842nm}_{\sigma}&=n_{\rm 2p_8,0}A_{01}\gamma_{0,1}(n_{\rm 1s_4,1})+n_{\rm 2p_8,1}A_{10}\gamma_{1,0}(n_{\rm 1s_4,0})+n_{\rm 2p_8,2}A_{21}\gamma_{2,1}(n_{\rm 1s_4,1})\\
&+n_{\rm 2p_8,0}A_{0-1}\gamma_{0,-1}(n_{\rm 1s_4,-1})+n_{\rm 2p_8,-1}A_{-10}\gamma_{-1,0}(n_{\rm 1s_4,0})\\&+n_{\rm 2p_8,-2}A_{-2-1}\gamma_{-2,-1}(n_{\rm 1s_4,-1})
\end{split}
\end{equation}
and for 801 nm branch:
\begin{equation}
\begin{split}
I^{\rm 801 nm}_{\pi}&=n_{\rm 2p_8,1}A_{11}\gamma_{1,1}(n_{\rm 1s_5,1})+n_{2p_8,2}A_{22}\gamma_{2,2}(n_{\rm 1s_5,2})\\
&+n_{\rm 2p_8,-1}A_{-1-1}\gamma_{-1,-1}(n_{\rm 1s_5,-1})
+n_{2p_8,-2}A_{-2-2}\gamma_{-2,-2}(n_{\rm 1s_5,-2})\\
\\
I^{\rm 801 nm}_{\sigma}&=2n_{\rm 2p_8,0}A_{01}\gamma_{0,1}(n_{\rm 1s_5,1})+2n_{\rm 2p_8,1}A_{10}\gamma_{1,0}(n_{\rm 1s_5,0})\\
&+2n_{\rm 2p_8,2}A_{21}\gamma_{2,1}(n_{\rm 1s_5,1})+2n_{\rm 2p_8,1}A_{12}\gamma_{1,2}(n_{\rm 1s_5,2})\\
&+n_{\rm 2p_8,0}A_{0-1}\gamma_{0,-1}(n_{\rm 1s_5,-1})+n_{\rm 2p_8,-1}A_{-10}\gamma_{-1,0}(n_{\rm 1s_5,0})\\
&+n_{\rm 2p_8,-2}A_{-2-1}\gamma_{-2,-1}(n_{\rm 1s_5,-1})+n_{\rm 2p_8,-1}A_{-1-2}\gamma_{-1,-2}(n_{\rm 1s_5,-2})
\end{split}
\end{equation}
\\
Similar sets of equations can be constructed for each pumped transition resulting in $\pi$ and $\sigma$ radiation rates for both radiation branches 842 and 801 nm as observed in the measurements. 
With the proposed description, it is evident that pumping of a same state through different transitions, as $A_{11}$ and $A_{10}$, would lead to the identical polarization although with different intensities determined mostly by the laser pumping rate through the targeted 1s state densities ($\rm n_{1s4,1}$ and $\rm n_{1s4,0}$ respectively).  

Equations constructed for each pumped transition ($A_{00},A_{11},A_{21},A_{10}$ and $A_{0-1}$) will effectively create 3 equations describing $\pi$ to $\sigma$ ratios at each radiation branch resulting in 6 coupled equations with 6 variables ($n_{{\rm 1s_4},m_j=0}$, $n_{{\rm 1s_4},m_j=1}$, $n_{{\rm 1s_5},m_j=0}$, $n_{{\rm 1s_5},m_j=1}$, $n_{{\rm 1s_5},m_j=2}$ and $k_{\rm dis}$) assuming symmetry in the self-alignment of 1s states ($n_{-m_j}=n_{m_j}$). The variables can be reduced to 3 ($n_{{\rm 1s_4},m_j=0}$, $n_{{\rm 1s_5},m_j=0}$ and $k_{\rm dis}$) because 1s sub-level densities belonging to the same state have a constant relation to each other. This is confirmed within this study and by the previous work \cite{Bergert_2019} as also the symmetry by LAS. The system of equations is mostly coupled through the self-absorption coefficients. The system of equations was solved with 3 variables using least-square method by minimizing the error between the measured and modeled rations of polarities.

\begin{figure}[h]
	\centering
	\includegraphics[width=0.5\linewidth]{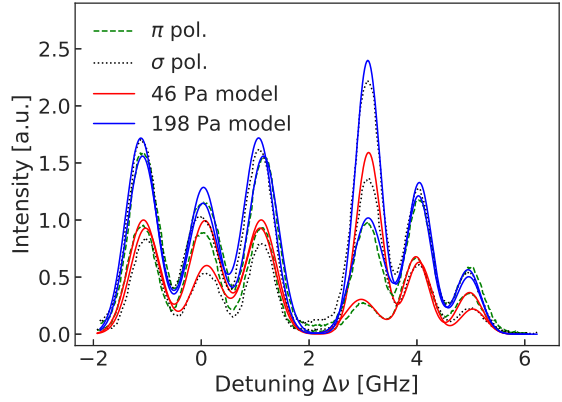}
	\caption{Modeled and measured fluorescence at 842 nm for two different pressures and both polarization components.}
	\label{fig:s4model}
\end{figure}

\begin{figure}[h]
	\centering
	\includegraphics[width=0.5\linewidth]{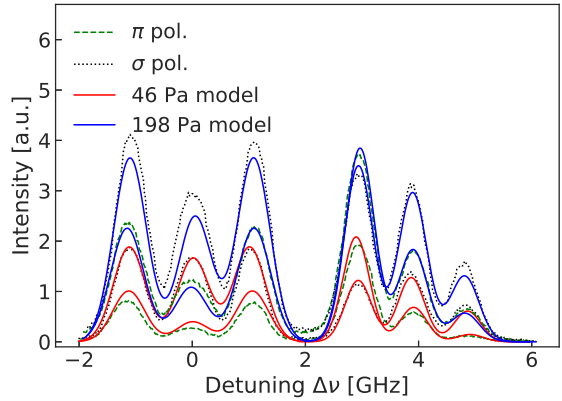}
	\caption{Modeled and measured fluorescence at 801 nm for two different pressures and both polarization components.}
	\label{fig:s5model}
\end{figure}

Solution of the system of equations will result in numerical evaluation of $\rm 1s_4$ and $\rm 1s_5$ multiplet densities and effective disalignment rate in form $c_{\rm dis}$=$Nk_{\rm dis}$. Measured fluorescence and modeled intensities at two different pressures (red low and blue medium pressure) are presented in figure \ref{fig:s4model} for 842 nm and figure \ref{fig:s5model} for 801 nm branch. The quality of the fits is clearly visible from the presented example where besides the reconstructed profiles, relative changes in the intensities between two different pressures could also be reliably reproduced. Similar quality of the fits is achieved for all measured pressure settings. In most cases the deviation is not larger than few percent making this analysis highly efficient. Here it should be emphasized that since only the ratios between the two polarities for each branch were considered in evaluation, there is no need for intensity calibration of the detection system for the two observed wavelengths.

\begin{figure}[h]
	\centering
	\includegraphics[width=0.5\linewidth]{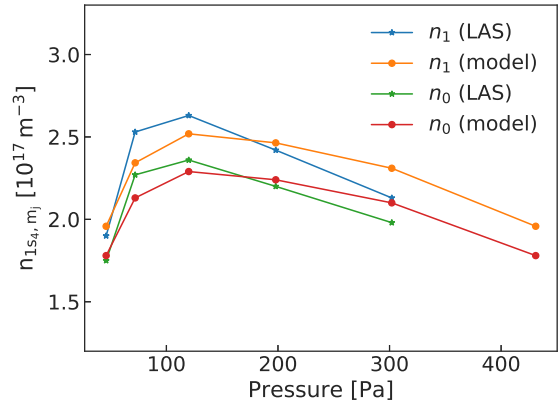}
	\caption{Density of $\rm 1s_4$ sub-levels measured by LAS and evaluated from TDLIF under different pressures ($\propto$ disalignment rate $c_{dis}$).}
	\label{fig:s4density}
\end{figure}

\begin{figure}[h]
	\centering
	\includegraphics[width=0.5\linewidth]{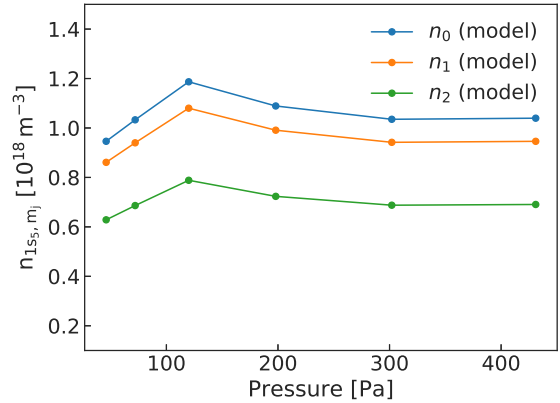}
	\caption{Densities of $\rm 1s_5$ sub-levels evaluated from TDLIF measurements under different pressures ($\propto$ disalignment rate $c_{dis}$).}
	\label{fig:s5density}
\end{figure}

Reconstructed $\rm 1s_4$ and $\rm 1s_5$ multiplets densities are presented in figures \ref{fig:s4density} and \ref{fig:s5density} respectively, for series of different pressures. The estimated $\rm 1s_4$ densities were further compared with the results from direct laser absorption measurements. The estimated densities by this two independent methods are in close agreement accounting for the complexity of the measurements and the evaluation. Alignment between the magnetic sub-levels has been also successfully reconstructed for both states. Reconstructed $\rm 1s_5$ multiplet densities can be indirectly compared with the results from the LAS measurements presented in \cite{Bergert_2019} where similar densities and identical negative alignment was observed under very similar conditions.

\begin{figure}[h]
	\centering
	\includegraphics[width=0.5\linewidth]{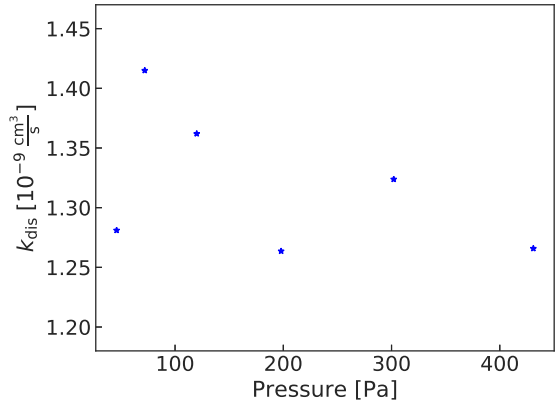}
	\caption{Estimation of disalignment rate coefficient at different measured pressure.}
	\label{fig:druck}
\end{figure}

The pressure at the measurement point (between the magnets) evaluated for used gas flow conditions was estimated to be in the range between 46 and 431 Pa. The estimated pressures by LIF were in between the values measured at the extraction side (lower pressure) and in gas flow line (higher pressure), closer to the high pressure side. Based on the pressure estimations, the disalignment rate coefficient $k_{dis}$ was estimated to be in the range from 1.26 to 1.41 $\times 10^{-9}\, \rm{\frac{cm^3}{s}}$ as presented in figure \ref{fig:druck}. 

The estimated values at different pressures are in a good agreement with the reported value of 1.45 $\times 10^{-9}\, \rm{\frac{cm^3}{s}}$ in \cite{grandin1973sections} which is evaluated at higher temperature of 380 K and by a different method. Evaluated temperature of 340 K by LAS in our conditions is lower which could reduce the disalignment coefficient according to the temperature dependency reported in \cite{Grandin1981}. The deviation in estimated rate coefficient at different pressures could be the result of poor estimation of pressure inside the tube based on measurements of relatively small shifts (of about 50 MHz) and robust model for pressure estimation. 

The disalignment rate coefficients for most of argon $\rm 2p$ states (in range from 1 -- 4 $\times 10^7\, \rm{s^{-1}}$) reported in \cite{grandin1973sections} are in excellent agreement with our estimated rate coefficients for $\rm 2p_8$ in range of 1 -- 12 $\times 10^7\, \rm{s^{-1}}$ between 46 -- 431 Pa, rendering disalignment effect highly important for accurate description of the 2p state densities in a magnetized plasma. 

\begin{figure}[h]
	\centering
	\includegraphics[width=0.5\linewidth]{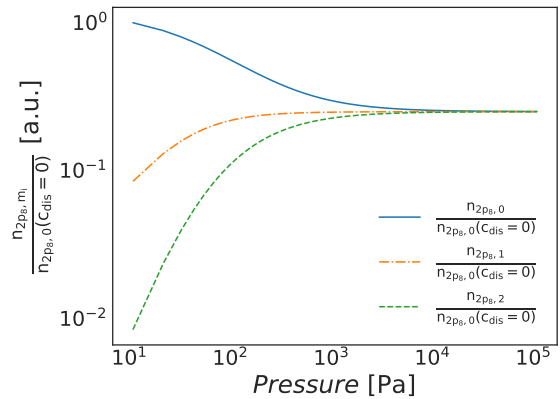}
	\caption{Normalized $\rm 2p_8$ sub-level densities solved for pumping $\rm m_j=0$ to $\rm m_i=0$ transition. In this particular case $m_i=\pm2,\pm1$ is same so that $m_i=-2,-1$ are not shown.}
	\label{fig:2p8}
\end{figure}
The influence of pressure on sub-state densities in the presence of laser pumping is illustrated in figure \ref{fig:2p8} where steady state solution for pumping of $m_j=0$ state by $\rm A_{00}^{842\, nm}$ transition is shown. The pressure dependent evolution of densities were calculated using constant and low values of $\rm 1s_4$ and $\rm 1s_5$ sub-state densities normalized by the value at $\rm c_{\rm dis}=0$. It is evident that increase of pressure would lead to increase of the population of other not-pumped sub-levels effectively changing polarization and overall intensity of the fluorescence signals. Since the central ($A_{00}$) transition is pumped the resulting alignment will be symmetric while pumping of other states $m_j=\pm1,\pm2$ would result in strong asymmetric alignment. 

The increase of the 2p sub-level densities which are not pumped is a result of a more efficient mixing between the sub-levels until the point when densities within the multiplet are equally distributed having no alignment. 

\section{Conclusion}
A method to evaluate TDLIF measurements accounting for intra- and inter-multiplet transitions has been proposed resulting in full quantitative evaluation of 1s states multiplets involved in the LIF scheme and evaluation of the disalignment coefficient for argon. The method was described for the $\rm 2p_8$ to $\rm 1s_4$ and $\rm 1s_5$ schemes but can be easily adopted for other 2p argon states.
The evaluated densities were in good agreement with the values measured by LAS for wide pressure range. The disalignment rate constant evaluated within same analysis showed good agreement compared with restricted literature values. We also showed that the interpretation of LIF measurements in magnetized plasma conditions is strongly pressure dependent through the disalignment interactions.

Proposed measurement method and analysis could further improve an evaluation of a magnetized plasma conditions and provide information valuable for further development of diagnostics methods based purely on optical emission spectroscopy. With the systematic evaluation of disalignment for argon 2p levels and quantitative LIF done
on various pressures, the densities of all 1s states would be well-known, providing basis for further development of an appropriate line branching method for evaluation of electron densities and temperature. These properties are making LIF measurements in magnetized plasma as highly efficient tool for quantitative plasma diagnostics while providing basis for further understanding and description of light transport properties.

\ack This work is supported by the Deutsche Forschungsgemeinschaft (DFG).
\section*{References}
\bibliographystyle{iopart-num}
\bibliography{paper.bib}
\end{document}